\documentclass[preprint,authoryear,3p,twocolumn,12pt]{elsarticle}

\usepackage{euscript}
\usepackage{amsfonts}
\usepackage{graphicx}
\usepackage{amsmath}
\usepackage{amssymb}
\usepackage{euscript}
\usepackage{color}
\usepackage{epsfig}
\usepackage{latexsym}
\usepackage{dcolumn}
\usepackage{bm}
\usepackage{amsthm}
\usepackage{array,longtable}

\usepackage{calc}
\journal{New Astronomy}

\begin{document}

\begin{frontmatter}

\title{Observational Constraints on the Normal Branch of a Warped DGP Cosmology}


\author[label1]{Tahereh Azizi}
\ead{t.azizi@umz.ac.ir}
\author[label2,label3]{M. Sadegh Movahed}
\ead{m.s.movahed@ipm.ir}

\author[label1]{Kourosh Nozari}\ead{knozari@umz.ac.ir}

\address[label1]{Department of Physics, Faculty of Basic
Sciences, University of Mazandaran,\\
P. O. Box 47416-95447, Babolsar, IRAN\\}
 \address[label2]{Department of Physics, Shahid Beheshti University, G.C., Evin, Tehran 19839, Iran}

\address[label3]{School of Astronomy, Institute for Research in
Fundamental Sciences, (IPM), P. O. Box 19395-5531, Tehran, Iran}

\begin{abstract}
We investigate observational constraints on the normal branch of the
warped DGP braneworld cosmology by using observational data from
Type Ia Supernovae (SNIa), Baryon Acoustic Oscillations (BAO),
Cosmic Microwave Background (CMB) and Baryon Gas Mass Fraction of
cluster of galaxies. The best fit values of model free parameters
are: $\Omega_m=0.240^{+0.050}_{-0.130}$ and
$\Omega_{r_c}=0.000^{+0.014}$ at $1\sigma$ confidence interval by
using Gold sample SNIa$+$CMB shift parameter$+$BAO$+$Gas mass
fraction of baryons in cluster of galaxies. The results for essence
sample SNIa combined with CMB shift parameter, BAO and Baryon Gas
mass fraction correspont to: $\Omega_m= 0.220^{+0.020}_{-0.170}$ and
$\Omega_{r_c}=0.000^{+0.025}$ at $68.3\%$ confidence interval. We
determine the age of the universe by using these best fit values. We
also study the effective cosmological dynamics on the brane via an
effective equation of state parameter and the deceleration parameter
to conclude that an effective phantom-like behavior arises in this
scenario. \\


\end{abstract}


\end{frontmatter}

\section{Introduction}
The accelerated expansion of the universe supported by recent
observational data [1] is one of the most important discoveries in
the last decade for the cosmology community. Within the framework of
the general relativity, the acceleration could be associated with
the so-called \emph{dark energy}, whose theoretical nature and
origin are still unknown for theorists. Cosmological constant or
vacuum energy with an equation of state parameter $\omega=-1$, is
the most popular candidate for dark energy but unfortunately, it
suffers from some serious problems such as fine-tuning and
coincidence problems. Therefore, a number of models containing
dynamical dark energy have been proposed as the mechanism for
late-time cosmic speed up [2]. Some of them are quintessence,
k-essence, phantom scalar field, chaplygin gas models and so on.
Another alternative approach to explain the late-time cosmic speed
up is modification of the geometric sector of the Einstein field
equations leading to \emph{modified gravity} [3]. In the spirit of
modified gravity proposal, the Dvali-Gabadadze-Porrati (DGP)
braneworld scenario explains the late-time accelerated expansion in
its self-accelerating branch without need to introduce a dark energy
component on the brane. In this scenario our universe is a $3$-brane
embedded in a higher dimensional flat space-time (bulk). The
late-time acceleration is driven on large scales by leakage of
gravity from the brane into the bulk ( see for instance [4] and
references therein). On small scales, gravity is bound to the brane
and general relativity is recovered to a good approximation. Indeed,
the DGP model has two types of solutions (corresponding to two
possible embedding of brane in the bulk ): the self-accelerating,
($DGP^{(+)}$) branch and the normal, ($DGP^{(-)}$) branch. The
($DGP^{(+)}$) branch however, suffers from some instabilities such
as ghosts [5] and cannot describe the early stages of the universe
evolution properly. There are some extension of the DGP setup that
provide relatively wider parameter spaces with richer phenomenology.
One of these models is the \emph{warped DGP braneworld} (WDGP) [6]
which is a unified model of  Randall-Sundrum II (RSII) braneworld
scenario [7] and the DGP setup. In the RSII braneworld model,
gravity modifies in early ( high energy) epoches of the universe
evolution. The warped DGP scenario also gives a self-accelerating
phase in the brane cosmology. It is important to note that, the
self-accelerating branch gives an effective equation of state that
never can be less than $-1$ (always non-phantom behavior and
therefore no crossing of the phantom divide line). The other branch
of this scenario is the normal branch ($DGP^{(-)}$) which doesn't
self-accelerate but requires dark energy or modification of the
induced gravity on the brane to explain the late time acceleration.
It is possible to realize phantom-like effects (without phantom
matter) in this normal branch via screening of the brane
cosmological constant at late time [8].

This paper is devoted to explore observational status of the normal
branch of the warped DGP braneworld cosmology. Observational
constraints in DGP model with and without tension is investigated in
Ref. [9]. We impose constraints on the model parameters by using the
several recent observations such as distance measurements from type
Ia supernovae (SNIa) from the Gold [10] and Essence [11] surveys,
the baryon acoustic oscillations (BAO) measurement from the
large-scale correlation function of the Sloan Digital Sky Survey
(SDSS) [12], the position of the first peak of the cosmic microwave
background (CMB) from WMAP7 [13] and the baryon gas mass fraction of
cluster of galaxies [14]. The structure of the paper is as follows:
in section $2$, we introduce the model and its cosmological
implications. In sec. $3$ we explore the effect of the WDGP model on
the geometrical parameters of the universe. Section $4$ includes the
observational constraints on the model parameters space while in
section $5$ the detailed results are presented. In section $5$ we
perform a detailed comparison between age of the oldest objects of
the universe and the age result that obtained from the best fit
values of our model parameter space. We also study the effective
dynamics of the model in this section. Finally section $6$ is
devoted to the concluding remarks.
\section{The Model}
The action of the warped DGP braneworld model can be written as
follows [6,15]
\begin{equation}
{\cal{S}}={\cal{S}}_{bulk}+{\cal{S}}_{brane},
\end{equation}
$${\cal{S}}=\int_{bulk}d^{5}X\sqrt{-{}^{(5)}g}\bigg[\frac{1}{2\kappa_{5}^{2}}
{}^{(5)}R+{}^{(5)}{\cal{L}}_{m}\bigg]+$$
\begin{equation}
\int_{brane}d^{4}x\sqrt{-g}\bigg[\frac{1}{\kappa_{5}^{2}}
K^{\pm}+{\cal{L}}_{brane}(g_{\alpha\beta},\psi)\bigg].
\end{equation}
Here ${\cal{S}}_{bulk}$ is the action of the bulk,
${\cal{S}}_{brane}$ is the action of the brane and ${\cal{S}}$ is
the total action. $X^{A}$ with $A=0,1,2,3,5$ are coordinates in the
bulk, while $x^{\mu}$ with $\mu=0,1,2,3$ are induced coordinates on
the brane. $\kappa_{5}^{2}$ is 5-dimensional gravitational constant.
${}^{(5)}R$ and ${}^{(5)}{\cal{L}}_{m}$ are 5-dimensional Ricci
scalar and matter Lagrangian respectively. $K^{\pm}$ is trace of the
extrinsic curvature on either sides of the brane.
${\cal{L}}_{brane}(g_{\alpha\beta},\psi)$  is the effective
4-dimensional Lagrangian. The action ${\cal{S}}$ is actually a
combination of the Randall-Sundrum II and the DGP model. In other
words, an induced curvature term is appeared on the brane in the
Randall-Sundrum II model. Now we consider the brane Lagrangian as
follows
\begin{equation}
{\cal{L}}_{brane}(g_{\alpha\beta},\psi)=\frac{\mu^2}{2}R-\lambda+L_{m},
\end{equation}
where $\mu$ is a mass parameter, $R$ is the Ricci scalar of the
brane, $\lambda$ is the tension of the brane and $L_{m}$ is the
Lagrangian of the other matter fields localized on the brane. We
assume that bulk contains only a negative cosmological constant,
$\Lambda_{5}$. With these choices, action (1) gives either a
generalized DGP or a generalized RS II model: it gives DGP model if
$\lambda=0$ and $\Lambda_{5}=0$, and gives RS II model if $\mu=0$.
The generalized Friedmann equation on the brane is as follows [6]
\begin{equation}
H^{2}+\frac{k}{a^{2}}=\frac{1}{3\mu^2}\bigg[\rho+\rho_{0}\Big(1+\varepsilon
{\cal{A}}(\rho,a)\Big)\bigg],
\end{equation}
where $\varepsilon=\pm 1$ is corresponding to two possible branches
of solutions (two different embedding of the brane) in this warped
DGP model and
${\cal{A}}=\bigg[{\cal{A}}_{0}^{2}+\frac{2\eta}{\rho_{0}}
\Big(\rho-\mu^{2}\frac{{\cal{E}}_{0}}{a^{4}}\Big)\bigg]^{1/2}$ where
\,\, ${\cal{A}}_{0}\equiv
\bigg[1-2\eta\frac{\mu^{2}\Lambda_{5}}{\rho_{0}}\bigg]^{1/2}$,\,\,
$\eta \equiv\frac{6m_{5}^{6}}{\rho_{0}\mu^{2}}$\,\, with
$0<\eta\leq1$ \,\,and \,\,$\rho_{0}\equiv
m_{\lambda}^{4}+6\frac{m_{5}^{6}}{\mu^{2}}$.\, By definition,
$m_{\lambda}= \lambda^{1/4}$ and $m_{5}=k_{5}^{-2/3}$. \,Also,
${\cal{E}}_{0}$ is an integration constant and corresponding term in
the generalized Friedmann equation is called dark radiation term. We
neglect dark radiation term in what follows. In this case, the
generalized Friedmann equation (4) takes the following form
\begin{equation}
H^{2}+\frac{k}{a^2}=\frac{1}{3\mu^2}\bigg[\rho+\rho_{0}+\varepsilon
\rho_{0}\Big({\cal{A}}_{0}^{2}+\frac{2\eta\rho}{\rho_{0}}\Big)^{1/2}\bigg],
\end{equation}
where $\rho\equiv\rho_{m}$ is the energy density of dark matter on
the brane.

\subsection{Cosmological Implications} In this section we study
cosmological dynamics on the DGP brane embedded in a warped bulk
Manifold. To this end, we assume a flat FRW universe on the warped
DGP brane. In this setup we can rewrite equation (5) as follows
\begin{equation}
 H^{2}=\frac{\rho+\lambda}{3\mu^{2}}+\frac{1}{2r_{c}^{2}}\bigg[1\pm\sqrt{1+4r_{c}^{2}
 \Big(\frac{\rho+\lambda}{3\mu^{2}}-\frac{{}\Lambda_{5}}{3}\Big)}\bigg]
\end{equation}
where $r_{c}=\frac{m_{4}^{2}}{2m_{5}^{3}}$ is the DGP crossover
scale. In the distance scale lower than this scale, gravity behaves
as usual general relativistic one but in the distance scales higher
than the crossover scale, gravity leaks to the extra dimension and
this leakage leads to weakness of gravity in the large scales, so
the universe expansion accelerates. The upper sign in equation (6)
corresponds to the self-accelerating branch of the model. Taking the
lower sign of this equation results a very interesting feature.
Indeed, if we assume a model universe with standard cold dark matter
(SCDM) with $\rho_{m}=\rho_{0_m} \Big(\frac{a_{0}}{a}\Big)^{3}$ the
accelerating behavior of the model can be recovered by  rewriting
Friedmann equation (6) as follows
\begin{equation}
 H^{2}=\frac{\rho_{0_m}a_{0}^{3}}{3\mu^{2}a^{3}}+\Lambda_{eff}\,,
\end{equation}
where $\Lambda_{eff}$ mimics the role of an effective cosmological
constant on the brane (note that it is not actually a constant!) and
it can be decomposed into two parts as follows
$$\Lambda_{eff}=\bigg(\frac{\lambda}{3\mu^{2}}+\frac{1}{2r_{c}^{2}}\bigg)-$$
\begin{equation}
\frac{1}{2r_{c}^{2}}\sqrt{1+4r_{c}^{2}\bigg(\frac{\rho_{0}a_{0}^{3}}{3\mu^{2}a^{3}}+
\frac{\lambda}{3\mu^{2}}-\frac{{}\Lambda_{5}}{3}\bigg)}
\end{equation}
The first two terms appeared in parenthesis on the right hand side
of this relation could be considered collectively as a cosmological
constant term on the brane and the last term on the right hand side
(the square root) screens the effect of the brane cosmological
constant in the same way as has been pointed out by Lue and Starkman
[8]. In this situation, the effective cosmological constant
$\Lambda_{eff}$ on the brane increases with time due to dynamical
screening effect, that is, reduction of the second term on the right
hand side of (8) with cosmic time. In fact the normal branch of the
model has the key property that brane is extrinsically curved so
that shortcuts through the bulk allow gravity to screen the effects
of the brane energy-momentum contents at Hubble parameters $H\sim
r_{c}^{-1}$. The screening effect is a result of leakage of gravity
to the extra dimension at late times.\\
For future purposes, it is useful to express the Friedmann equation
(6) in a dimensionless form as follows
 $$E^{2}(z)=\frac{ H^{2}(z)}{
 H_{0}^{2}}=\Omega_{m}(1+z)^{3}+\Omega_{\lambda}+2\Omega_{r_{c}}-$$
 \begin{equation}
 2\sqrt{\Omega_{r_{c}}}\sqrt{\Omega_{m}(1+z)^{3}+\Omega_{\lambda}+
 \Omega_{r_{c}}+\Omega_{\Lambda_{5}}}
\end{equation}
where $\Omega_{m}=\frac{\rho_{0m}}{3\mu^{2}H_{0}^{2}}$\, ,\,
$\Omega_{\lambda}=\frac{\lambda}{3\mu^{2}H_{0}^{2}}$\, ,\,
$\Omega_{r_{c}}=\frac{1}{4r_{c}^{2}H_{0}^{2}}$\,
$\Omega_{{}\Lambda_{5}}=\frac{-\Lambda_{5}}{3H_{0}^{2}}$ and
$H_0=100h\quad km/s/Mpc$. Taking $z=0$ imposes a constraints on the
model parameters as follows
\begin{footnotesize}
\begin{equation}
\Omega_{m}+\Omega_{\lambda}+2\Omega_{r_{c}}-
 2\sqrt{\Omega_{r_{c}}}\sqrt{\Omega_{m}+\Omega_{\lambda}+
 \Omega_{r_{c}}+\Omega_{{}\Lambda_{5}}}=1\,.
\end{equation}
\end{footnotesize}
The general relativistic limit can be recovered if we set
$\Omega_{r_{c}}=0$ (or $m_{5}=0$). In this case equations (10)
implies that $\Omega_{m}+\Omega_{\lambda}=1$. Considering the
tension of the brane as a cosmological constant this case leads to
the $\Lambda$CDM cosmology.

\section{The effect of WDGP on the Geometrical Parameters of the Universe}
The cosmological observations are mainly dependent on the background
geometry ( especially background spatial curvature) of the universe.
So, in this section we study the effect of the WDGP model on the
geometrical parameters of the universe. \\

\textbf{1. Comoving radial Distance}\\

One of the basic parameters in cosmology is the comoving radial distance.
For an object with redshift $z$ in a FRW background, this parameter
can be expressed as follows
\begin{footnotesize}
\begin{equation}
r(z)=\frac{c}{H_{0}\sqrt{|\Omega_{K}|}}{\cal{F}}\Big(\sqrt{|\Omega_{K}|}\int_{0}^{z}\frac{dz'}{H(z')/H_{0}}\Big)\,,
\end{equation}
\end{footnotesize} where ${\cal{F}}\equiv(x,\,\sin x,\,\sinh x)$\,
for $K=(0,\,1,\,-1)$\, respectively.\, $K$ marks curvature of the
spatial geometry and $\Omega_{K}= \frac{K}{3\mu^{2}H_{0}^{2}}$.
Figure $1$ shows the radial comoving distance versus the redshift
for different values of $\Omega_{{r}_{c}}$ in a flat background.
Clearly, increasing the values of $\Omega_{r_{c}}$ results in a
longer comoving distance. The mentioned quantity is a useful
quantity in the analysis of the luminosity distances of Supernova
type Ia.\
\begin{figure}[htp]
\begin{center}\includegraphics{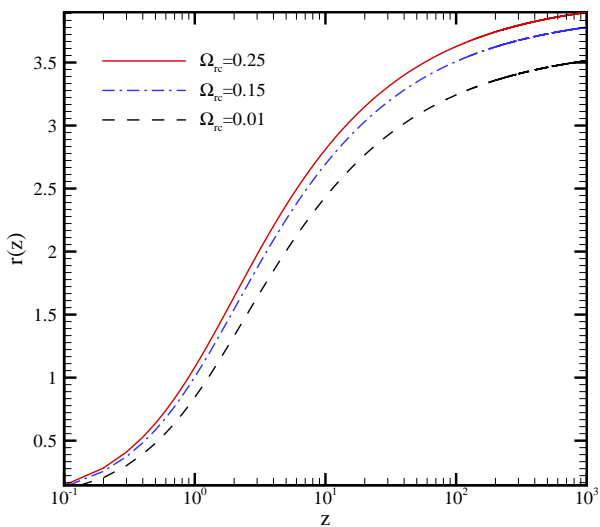} \vspace{6.5cm}
\end{center}
 \caption{\small {Comoving radial distance versus the redshift for different
 values of $\Omega_{{r}_{c}}$ in a flat FRW universe. The unit of vertical axis is $c/H_0$.}}
\end{figure}

\textbf{2. Angular size}\\

The apparent angular size of an object located at the cosmological
distance is another important parameter that can be affected by the
cosmological model during the history of the universe. If we take
the object to lie perpendicular to the line of sight and to have
physical extent $D$, the apparent angular size $\theta$ is given by
\begin{equation}
\theta=\frac{D}{d_{A}(z)}
\end{equation}
where $d_{A}(z)=r(z)/(1+z)$ is the angular diameter distance which
is a measure of how large objects appear to be in the universe. A
key application of equation (12) is in the study of features of the
cosmic microwave background radiation. The variation of apparent
angular size $\Delta\theta$ in terms of the $\Delta z$ is given by
\begin{equation}
\frac{\Delta z}{\Delta\theta}=H(z)r(z).
\end{equation}
This relation is the so-called Alcock-Paczynski test.
The advantage of the Alcock-Paczynski test is that in this case, instead of using a
standard candle, we use a standard ruler such as the
baryonic acoustic oscillation.
Figure $2$ shows $\Delta z/\Delta\theta$ for different values of
$\Omega_{{r}_{c}}$ in a flat FRW background. As the figure shows,
increasing of $\Omega_{{r}_{c}}$ results in larger values of $\Delta
z/\Delta\theta$.\\

\begin{figure}[htp]
\begin{center}\includegraphics{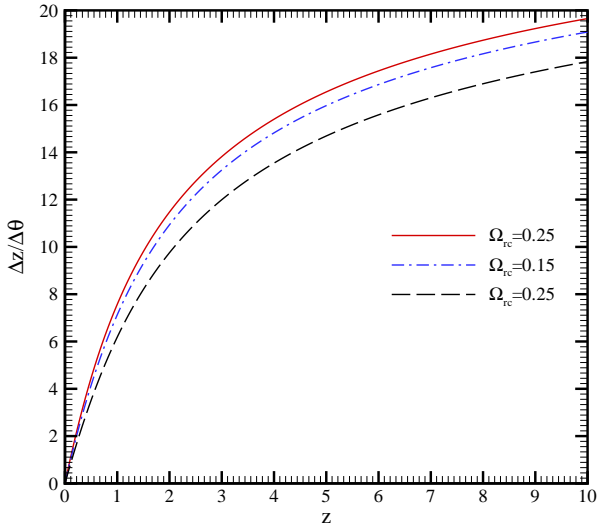} \vspace{6cm}
\end{center}
 \caption{\small {$\Delta z/\Delta\theta$ versus the redshift for different
 values of $\Omega_{{r}_{c}}$ in a flat FRW background.}}
\end{figure}

\textbf{3. Comoving Volume Element}\\

The comoving volume element is another geometrical parameter which
is used in number-count tests such as lensed quasars, galaxies, or
clusters of galaxies. The comoving volume element can be expressed
in terms of comoving distance and Hubble parameter as follows
\begin{equation}
f=\frac{dV}{dzd\Omega}=\frac{r^{2}(z)}{H(z)}\,.
\end{equation}
Figure $3$ shows the comoving volume element versus the redshift for
different values of $\Omega_{{r}_{c}}$ in a flat FRW background.
This figure indicates that the value of the comoving volume element
increases with the increasing of $\Omega_{{r}_{c}}$. We note the
quantities displayed in figures 1-3 are specified just by giving the
value of $\Omega_{{r}_{c}}$ alone. This is because we are interested
in the DGP character of the model. Nevertheless, in plotting these
figures we have used the values $\Omega_{m}=0.24$,
$\Omega_{{\Lambda}_{5}}=1$ and the other quantity $\Omega_{\lambda}$
is obtained via constraint equation, (10).
\begin{figure}[htp]
\begin{center}\includegraphics{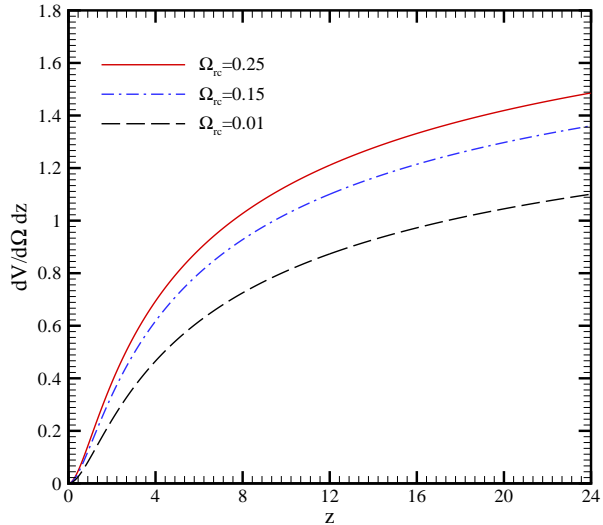} \vspace{6.5cm}
\end{center}
 \caption{\small {Comoving volume element versus the redshift for different
 values of $\Omega_{{r}_{c}}$ in a flat FRW background.}}
\end{figure}

\section{Observational Constraints}
In this section, we constrain the model parameters of the warped DGP
scenario by using the most recent observational data including the
SNIa data measurements as given by the gold and essence samples,
combined with the information from the BAO measurements by SDSS, the
CMB shift
parameter from WMAP7 observations and the baryon gas mass fraction.\\

\textbf{A: SNIa data}\\

Here we use some new sample supernova type Ia data such as the gold
sample compiled in [10] and essence sample compiled in [11] to
constrain the free parameters of the model. This observation
directly measures the apparent magnitude $m$ of a supernova versus
its redshift $z$. We note that deference between SNIa samples is in
their systematic errors and the range of redshift which the apparent
magnitude is determined .

The theoretical distance modulus is defined as
\begin{footnotesize}
\begin{equation}
\mu=m-M=5\log_{10}d_{L}+5\log_{10}\left(\frac{c/H_{0}}{1\quad
Mpc}\right)+25,
\end{equation}
\end{footnotesize}
where $M$ is the absolute magnitude that is believed to be constant
for all Type Ia supernovae and $d_{L}$ is the luminosity distance
that in general case can be expressed as follows
\begin{footnotesize}
\begin{equation}
d_{L}(z;\{\theta_i\})=\frac{(1+z)}{\sqrt{|\Omega_{K}|}}{\cal{F}}\left(\sqrt{|\Omega_{K}|}\int_{0}^{z}\frac{dz'}{E(z';\{\theta_{i}\})}\right)
\end{equation}
\end{footnotesize}
 where $\{\theta_{i}\}$ denotes the model
parameters and $E(z)$ is given by equation (9). We estimate the best
fit of the set of parameters $\{\theta_{i}\}$ by using a $\chi^{2}$
statistics, with
\begin{footnotesize}
\begin{equation}
\chi^{2}_{SN}(\{\theta_{i}\})=\sum_{j=1}^{N}\frac{[\mu_{obs}(z_{j};\{\theta_{i}\})
-\mu_{th}(z_{j};\{\theta_{i}\})]^{2}}{\sigma_{j}^{2}},
\end{equation}
\end{footnotesize}
In this relation, $N$ is the number of SNIa data points which is
different in several data sets, $\mu_{obs}$ is the observed distance
modulus and the $\sigma_{j}$ is the uncertainty in the observed
distance modulus, which is assumed to be Gaussian and uncorrelated
so that the likelihood is proportional to \, $\exp(-\chi^2/2)$. The
parameter $\overline{M}=5\log_{10}\big(\frac{c/H_{0}}{1\quad
Mpc}\big)+25$\ is a nuisance parameter and it is independent of the
data and the data sets. Following the techniques described in Ref.
[16], the minimization with respect to $\overline{M}$ can be made by
expanding the $\chi^{2}_{SN}$ of equation (17) with respect to
$\overline{M}$ as
\begin{equation}
\chi^{2}_{SN}(\{\theta_{i}\})=\tilde{A}-2M\tilde{B}+M^{2}\tilde{C}
\end{equation}
where\begin{scriptsize}
\begin{eqnarray}
\tilde{A}(\{\theta_{i}\})=\sum_{j=1}^{N}\frac{[\mu_{obs}(z_{j};\{\theta_{i}\})-
\mu_{th}(z_{j};\{\theta_{i}\},M=0)]^{2}}{\sigma_{j}^{2}},
\end{eqnarray}
\end{scriptsize}
and
\begin{scriptsize}
$$\tilde{B}(\{\theta_{i}\})=\sum_{j=1}^{N}\frac{\mu_{obs}(z_{j};\{\theta_{i}\})-
\mu_{th}(z_{j};\{\theta_{i}\},M=0)}{\sigma_{j}^{2}}$$

\begin{equation}
,\quad\quad\tilde{C}=\sum_{j=1}^{N}\frac{1}{\sigma_{j}^{2}}.
\end{equation}
\end{scriptsize}
Equation (18) has a minimum for $M=\frac{\tilde{B}}{\tilde{C}}$ \,
at\, $
\tilde{\chi}^{2}_{SN}(\{\theta_{i}\})=\tilde{A}(\{\theta_{i}\})-\frac{\tilde{B}^{2}(\{\theta_{i}\})}{\tilde{C}}$.
Using this equation the best fit values of model parameters as the
values that minimize $\chi^{2}_{SNIa}(\{\theta_{i}\})$ can be
obtained.

For the likelihood analysis we marginalize the likelihood function
$L \sim\exp(-\chi^{2}/2)$ over $h$. We adopted Gaussian priors such
that $h=0.705$ from the WMAP7 [13]. Table $1$ summarizes these priors. \\

\begin{table}
\begin{center}
\caption{Priors on the parameter space used in the likelihood
analysis.} \vspace{0.5 cm}
\begin{tabular}{|c|c|c|c|c|c|c|c|c|c|}
  \hline
  \hline Parameter&\,\,\,Prior\\
  \hline $\Omega_{m}$&  0.000\, - \, 1.000 \quad Top Hat\\
  \hline $\Omega_{r_{c}}$&  0.000\, - \, 1.000 \quad Top Hat\\
  \hline $\Omega_{\Lambda_{5}}$& -1.000\, -\, 1.000 \quad Top Hat \\
  \hline $h$&- \quad \quad - \\

   \hline
\end{tabular}
\end{center}
\end{table}

\textbf{B: CMB shift parameter}\\

We use the CMB data from WMAP7 observation that includes the shift
parameter $\cal{R}$ and the redshift of the decoupling epoch
$z_{*}$. The shift parameter $\cal{R}$ relates the angular diameter
distance to the last scattering surface, the comoving size of the
sound horizon at $z_{*}=1091.3$ and the angular scale of the first
acoustic peak in the CMB power spectrum of the temperature
fluctuations. The CMB shift parameter is approximated by (see for
more details Ref. [16])
\begin{equation}
{\cal{R}}=\frac{\sqrt{\Omega_{m}}H_{0}}{c}(1+z_*)d_A(z_{*}),
\end{equation}
where $d_A(z)$ is the angular diameter distance defined by equation
(12). The constraints on a typical model using CMB shift is obtained
from minimization of the quantity
\begin{equation}
\chi^{2}_{CMB}=\frac{[{\cal{R}}_{obs}-{\cal{R}}_{th}]^{2}}{\sigma_{CMB}^{2}},
\end{equation}
where ${\cal{R}}_{obs}=1.725$ is the observed value of the CMB shift
parameter performed from WMAP7 observation and its corresponding
$1\sigma$ error is $\sigma_{CMB}=0.018$ [13]. Also ${\cal{R}}_{th}$
corresponds to the theoretical value of shift parameter calculated
from equation
(21). \\

\textbf{C: BAO observation}\\

The baryonic acoustic oscillation (BAO) peak detected in the SDSS
luminous red Galaxies (LRG) is another tool to test the model
against observational data. BAO are described in terms of a
dimensionless parameter
\begin{scriptsize}
\begin{equation}
{\cal{A}}(z_{sdss};\{\theta_{i}\})=\sqrt{\Omega_{m}}\bigg[\frac{H_{0}d_{L}^{2}(z_{sdss};\{\theta_{i}\})}
{H(z_{sdss};\{\theta_{i}\})z_{sdss}^{2}(1+z_{sdss})^{2}}\bigg]^{1/3}\,.
\end{equation}
\end{scriptsize}
The $\chi^{2}$ for the BAO is given by
\begin{equation}
\chi^{2}_{sdss}=\frac{[{\cal{A}}_{obs}-{\cal{A}}_{th}]^{2}}{\sigma_{sdss}^{2}}.
\end{equation}
The observed value ${\cal{A}}_{obs}$ from the LRG is
${\cal{A}}_{obs}= 0.469\Big(\frac{n_{s}}{0.98}\Big)^{-0.35} \pm
0.017$ measured at $z_{sdss}=0.35$ [20]. Here $n_{s}=0.963$ is the
spectral index as measured by WMAP seven years observations [13].\\

\textbf{D: Gas mass fraction of cluster of galaxies}\\

Another cosmological test to constrain the parameters of the model
arises from baryon gas mass fraction of cluster of galaxies for a
range of redshifts
\begin{equation}
f_{gas}=\frac{M_{gas}}{M_{tot}}\,.
\end{equation}
The basic assumption underlying this method is that the baryon gas
mass fraction in clusters is constant, independent of redshift. This
method can give a constraint to the geometry of the universe with
the relation ${\cal{S}}_{gas}\propto d_{A}^{\frac{3}{2}}$ under the
assumption that this fraction should be approximately constant with
redshift. Following [17] (see also [16]), the $\chi^{2}$ expression
for gas mass fraction is given by
\begin{scriptsize}
\begin{equation}
\chi^{2}_{gas}(\{\theta_{i}\})=\sum_{j}\frac{[{\cal{S}}_{gas}^{obs}(z_{j};\{\theta_{i}\})-
{\cal{S}}_{gas}^{th}(z_{j};\{\theta_{i}\})]^{2}}{\sigma_{j}^{2}},
\end{equation}
\end{scriptsize}
where ${\cal{S}}_{gas}$ is a dimensionless parameter defined as
\begin{equation}
{\cal{S}}_{gas}=\frac{b}{1+\beta}\frac{\Omega_{b}}{\Omega_{m}}
\bigg(\frac{d_{A}^{flat}(z)}{d_{A}(z)}\bigg)^{\frac{3}{2}}.
\end{equation}
In this relation $d_{A}^{flat}$ is the angular diameter distance to
a cluster in the test model which is assumed to be SCDM (cold dark
matter) in this case, and $b$ is a bias factor suggesting that the
baryon fraction in clusters is slightly lower than for the universe
as a whole. Also $1+\beta$ is a factor taking into account the fact
that the total baryonic mass in clusters consists of both X-ray gas
and optically luminous baryonic mass (stars), the latter being
proportional to the former with proportionality constant
$\beta\simeq0.19\sqrt{h}$ [16-18]. The nuisance parameter
$\xi=\frac{b}{1+\beta}\frac{\Omega_{b}}{\Omega_{m}}$ should be
marginalized via expanding the $\chi^{2}_{gas}$ of equation (26)
with respect to $\xi$ which gives
\begin{equation}
\chi^{2}_{gas}=K-\frac{W^{2}}{Y}
\end{equation}
where
\begin{equation}
K=\sum_{j}\frac{{\cal{S}}_{gas}^{obs}(z_{j};\{\theta_{i}\})^{2}}{\sigma_{j}^{2}},
\end{equation}
and
$$W=\sum_{j}\frac{{\cal{S}}_{gas}^{obs}(z_{j};\{\theta_{i}\}).{\cal{S}}_{gas}^{th}
(z_{j};\{\theta_{i}\},\xi=1)}{\sigma_{j}^{2}},$$
\begin{equation}
Y=\sum_{j}\frac{{\cal{S}}_{gas}^{th}(z_{j};\{\theta_{i}\},\xi=1)^{2}}{\sigma_{j}^{2}}.
\end{equation}
We use the $26$ cluster data [14] for ${\cal{S}}_{gas}^{obs}$ to
obtain the best fit parameters of the model.

It is important to note that the above observational data are
uncorrelated since they are given by different experiments and
methods. So, we can construct a joint analysis as
\begin{equation}
\chi^{2}_{tot}=\chi^{2}_{SN}+\chi^{2}_{CMB}+\chi^{2}_{SDSS}+\chi^{2}_{gas}\,\,.
\end{equation}

\section{Results}
With these preliminaries, we have obtained the best fit parameters
of the normal branch of the WDGP model for SNIa data (gold and
essence datasets), the joint analysis of the SNIa and CMB, the
combined analysis of the SNIa, CMB and SDSS and finally the joint
analysis of the total datasets including cluster galaxies gas mass
fraction. Table $2$ shows the results of the observational
constraints on the free parameters of this model. Figure $4$ shows
the marginalized relative likelihood with respect to parameter
$\Omega_{m}$ and $\Omega_{r_{c}}$ fitted with SNIa gold sample,
SNIa+CMB,\, SNIa+CMB+SDSS experiments and SNIa+CMB+SDSS+gas mass
fraction observations. In figure $5$, we repeat these stages for
Essence sample of SNIa experiments together with other data sets. We
plot contour maps of $\Omega_{r_{c}}$ versus $\Omega_{m}$ in figure
$6$ for Gold Sample and combined observational data sets. In figure
$7$ we plot corresponding contour maps for Essence Sample and
combined observational data sets.
\newenvironment{cmdsyntax}
\addvspace{3.2ex plus 0.8ex minus 0.2ex}%
    \vskip -\parskip

\noindent
\begin{table}
\begin{center}
\begin{onecolumn}
\caption{The fitting results for WDGP model by using the SNIa (Gold
and Essence Samples), SNIa+CMB and SNIa+CMB+SDSS and
SNIa+CMB+SDSS+glaxies clusters gas mass fraction experiments in a
flat background.} \vspace{0.5 cm}
\end{onecolumn}
\noindent
\twocolumn
\begin{footnotesize}
\begin{tabular}{|l|c|c|c|c|c|c|c|c|c|}
  \hline
  \hline Observation &$\Omega_{m}$\,\,& $\Omega_{r_{c}}$\,\,& $\Omega_{\lambda}$\,\,
  &$\Omega_{\Lambda_{5}}$\,\,&$\chi^{2}_{min}/N_{d.o.f}$\,\,&Age(Gyr)\\
  \hline SNIa( Gold Sample)&$0.030^{+0.220}_{-0.025}$&$ 0.160^{+0.030}_{-0.150}$ &$0.168^{+0.360}_{-0.016}$& 0.009&0.923&
  $64.97^{+42.250}_{-35.855}$\\
  \hline SNIa(Gold)+CMB&$0.280^{+0.050}_{-0.130}$& $0.000^{+0.010}$ &$0.631^{+0.204}_{-0.117}$&0.939&0.943&$14.38^{+1.245}_{-3.344}$\\
 \hline SNIa(Gold)+CMB+SDSS&$0.260^{+0.060}_{-0.120}$&$ 0.000^{+0.012}$ &$0.644^{+0.302}_{-0.112}$&0.839&0.992&$14.51^{+1.448}_{-3.128}$ \\
 \hline SNIa(Gold)+CMB+SDSS+GAS&$0.240^{+0.050}_{-0.130}$& $0.000^{+0.014}$ &$0.660^{+0.174}_{-0.139}$&0.999&0.992&$14.85 ^{+1.561}_{-3.942}$\\

 \hline SNIa( Essence Sample)&$0.020^{+0.240}_{-0.015}$& $0.170^{+0.020}$ &$0.161^{+0.074}_{-0.037}$& 0.009&1.032&$78.38^{+56.018}_{-42.390}$\\
  \hline SNIa(Essence)+CMB&$0.230^{+0.020}_{-0.160}$& $0.000^{+0.019}$ &$0.691^{+0.133}_{-0.061}$&0.539&1.030&14.86$^{+1.435}_{-6.392}$\\
 \hline SNIa(Essence)+CMB+SDSS&$0.220^{+0.030}_{-0.190}$& $0.000^{+0.022}$ &$0.697^{+0.145}_{-0.095}$&0.689&1.054&15.07$^{+1.758}_{-6.977}$ \\
 \hline SNIa(Essence)+CMB+SDSS+GAS&$0.220^{+0.020}_{-0.170}$& $0.000^{+0.025}$ &$0.713^{+0.187}_{-0.125}$&0.119&1.044&$14.96^{+1.459}_{-7.888}$ \\
 \hline
\end{tabular}
\end{footnotesize}
\end{center}
\end{table}
\noindent
\begin{figure}[htp]
\begin{center}
\includegraphics{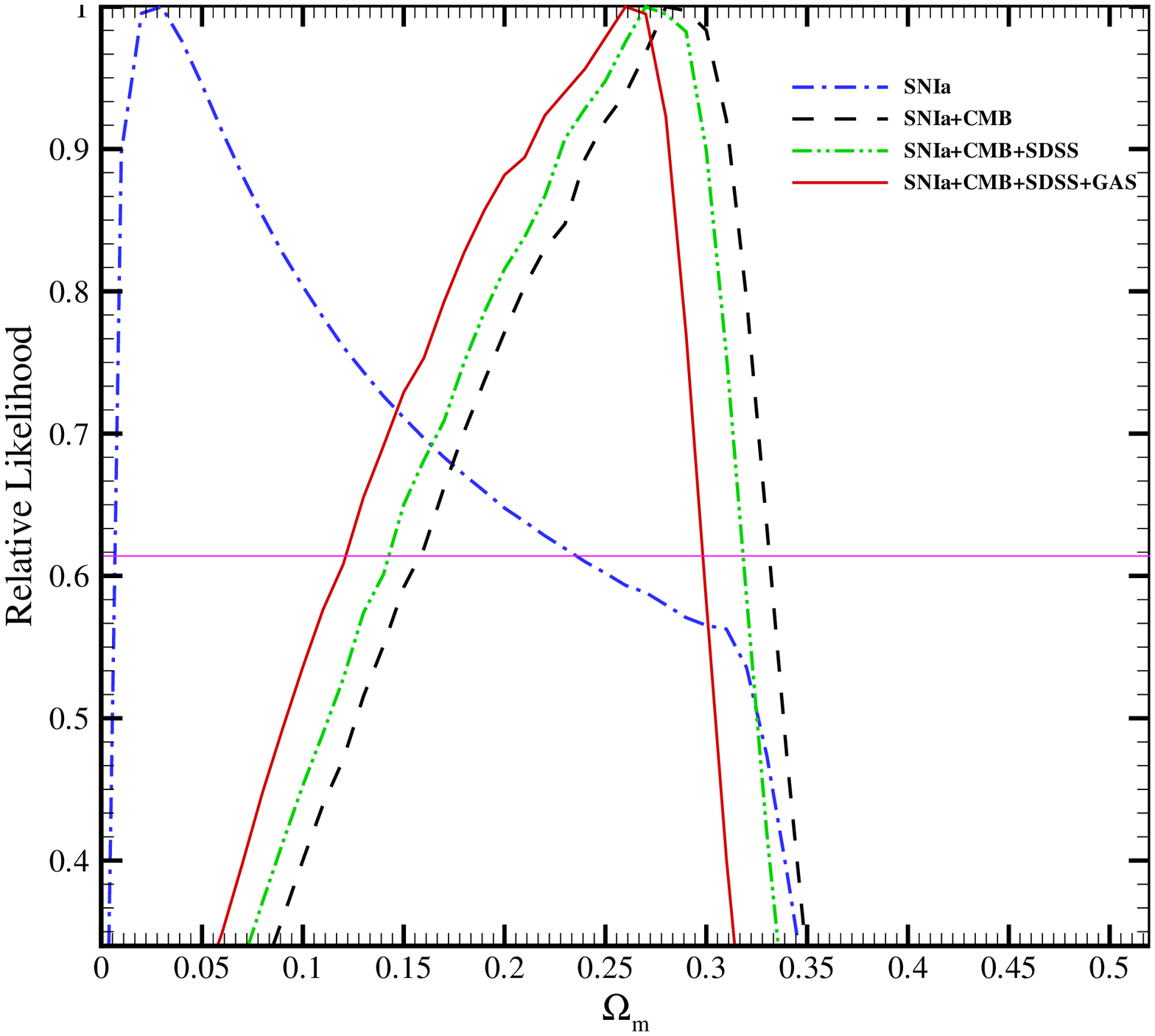} \vspace{2cm}\includegraphics{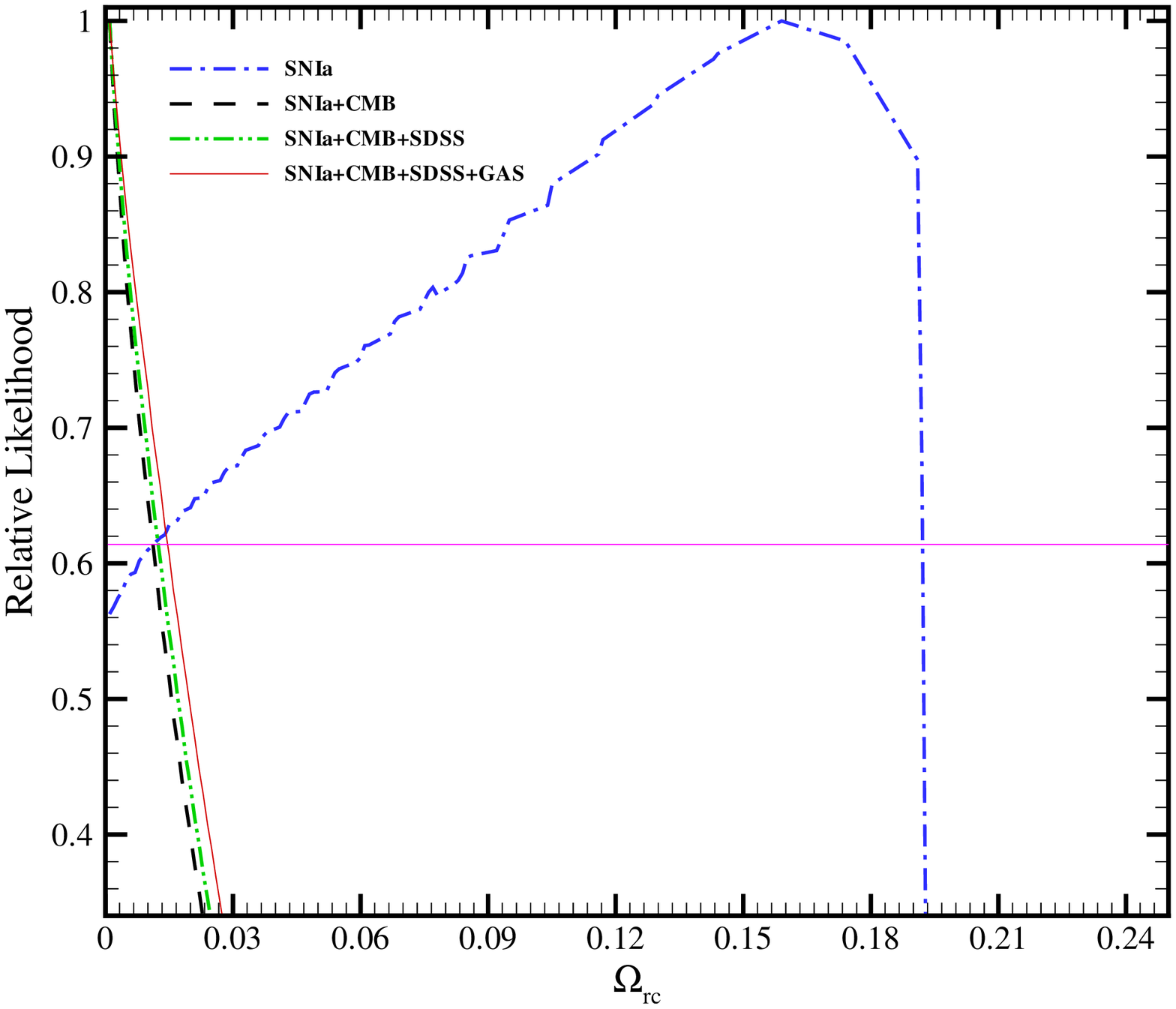}\vspace{4.5cm}
\begin{onecolumn}\caption{\small {Marginalized relative likelihood with respect to
parameter $\Omega_{m}$ (left) and $\Omega_{r_{c}}$ (right) fitted
with SNIa Gold Sample, SNIa (Gold)+CMB, SNIa(Gold)+CMB+SDSS and
SNIa(Gold)+CMB+SDSS+ galaxies clusters gas mass fraction
experiments. The intersection of each curve with horizontal solid
line is corresponding to the bound with $1\sigma$ confidence
level.}}\end{onecolumn}
\end{center}
\end{figure}
\noindent
\begin{figure}[htp]
\begin{center}
\includegraphics{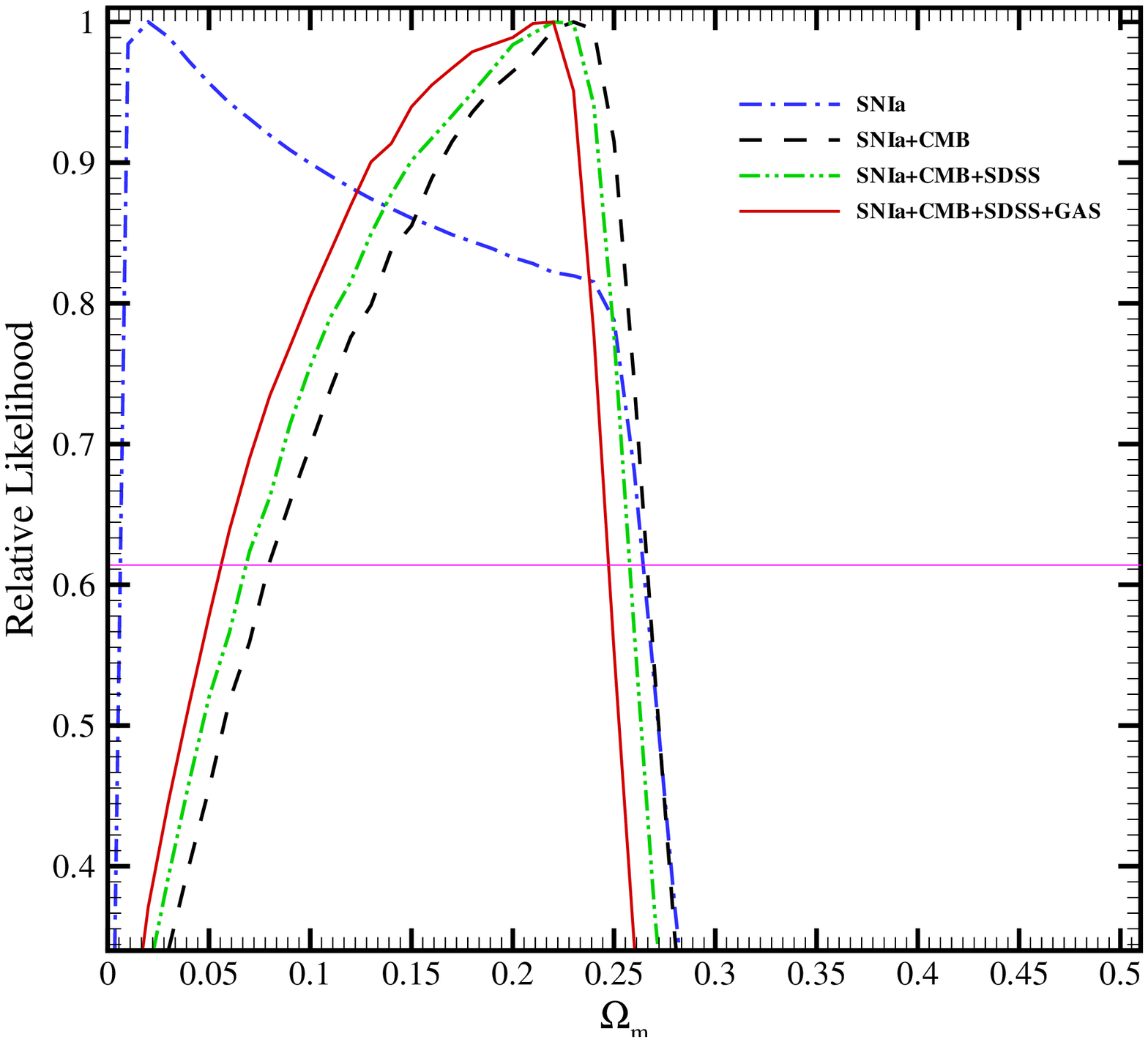} \vspace{2cm}\includegraphics{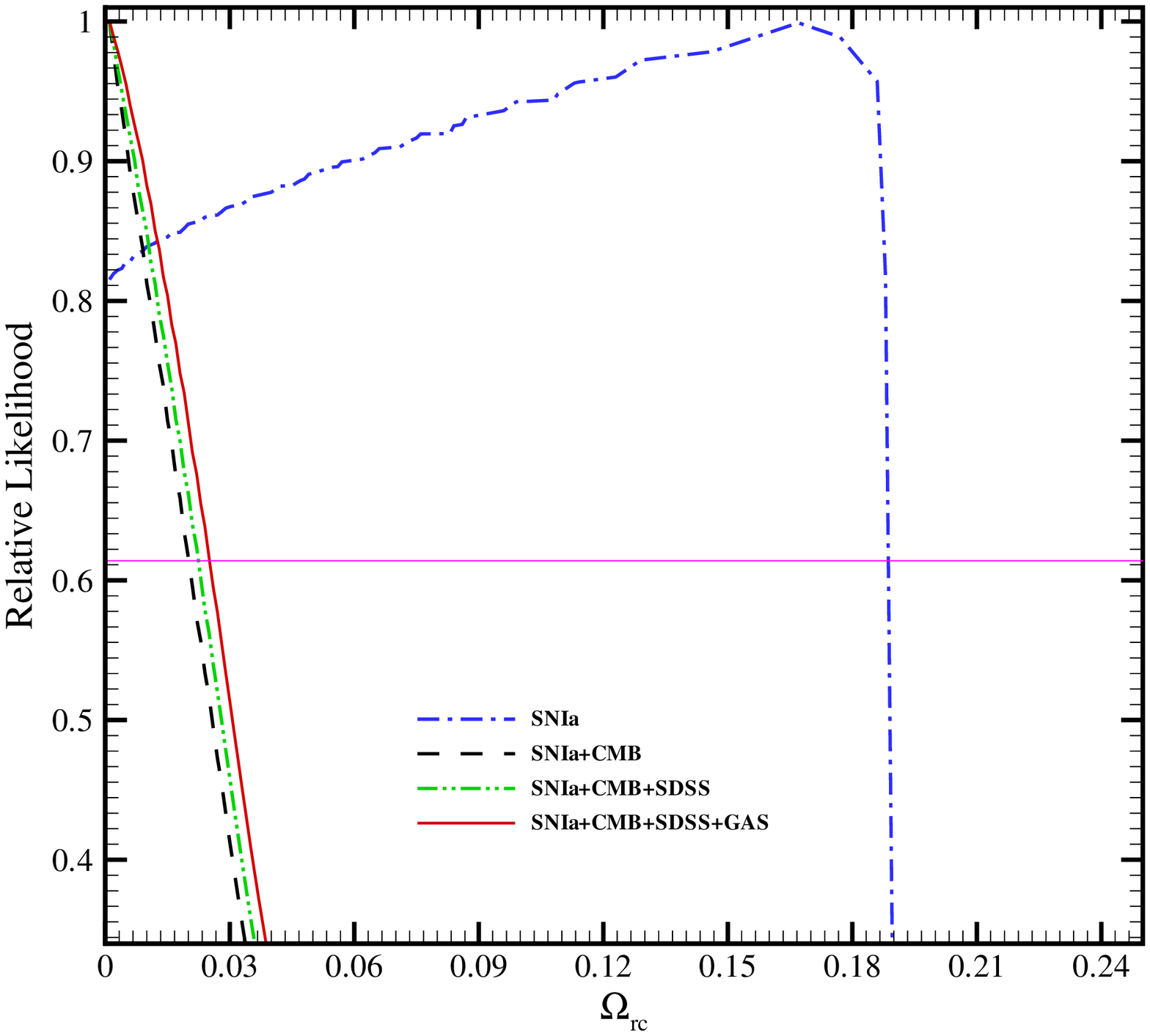}\vspace{4cm}
\caption{\small {\footnotesize{Marginalized relative likelihood with
respect to parameter $\Omega_{m}$ (left) and $\Omega_{r_{c}}$
(right) fitted with SNIa Essence sample, SNIa (Essence)+CMB, SNIa
(Essence)+CMB+SDSS and SNIa (Essence)+CMB+SDSS+galaxies clusters gas
mass fraction experiments. The intersection of each curve with
horizontal solid line is corresponding to the bound with $1\sigma$
confidence level.}}}
\end{center}
\end{figure}

\begin{figure}[htp]
\begin{center}\includegraphics{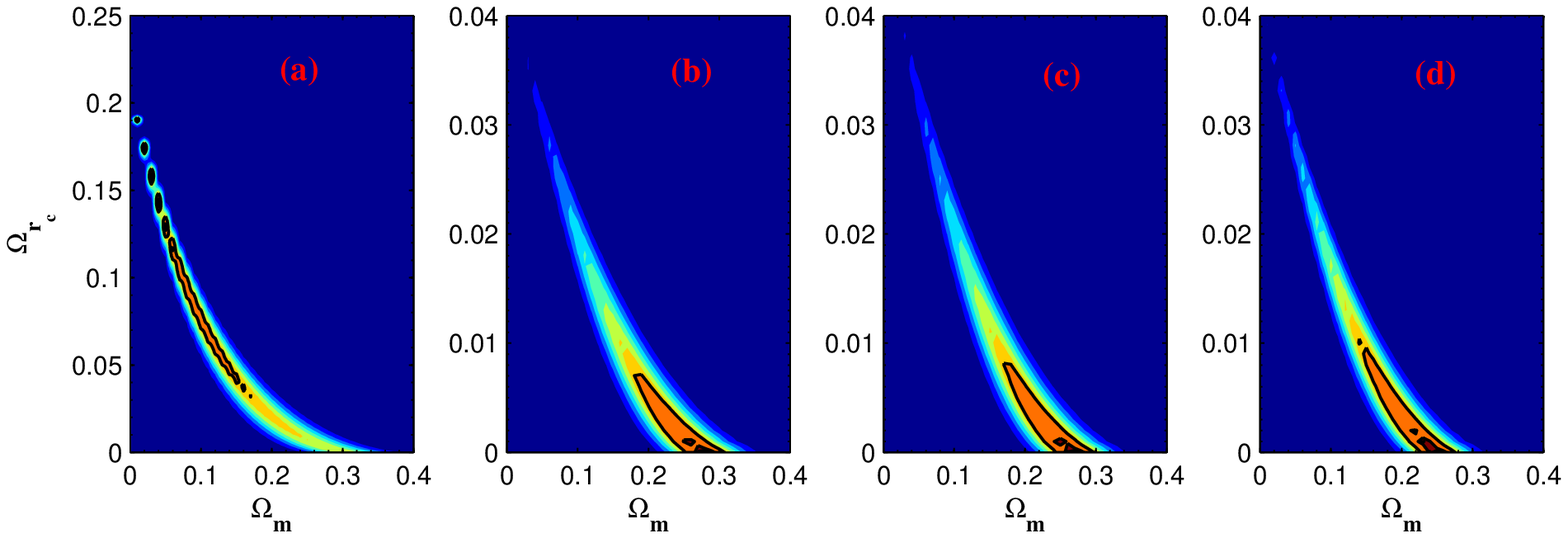} \vspace{3.5cm}
\end{center}
 \caption{\small {Contour maps of $\Omega_{r_{c}}$ versus $\Omega_{m}$
 for WDGP model with $1\sigma \,(68.3\%)$ and
$2\sigma \, (95.5\%)$ \footnotesize{confidence levels using SNIa
(a), SNIa+CMB (b), SNIa+CMB+SDSS (c) and SNIa+CMB+SDSS+Gas (d). We
used the Gold Sample for SNIa experiment.}}}
\end{figure}

\begin{figure}[htp]
\begin{center}\includegraphics{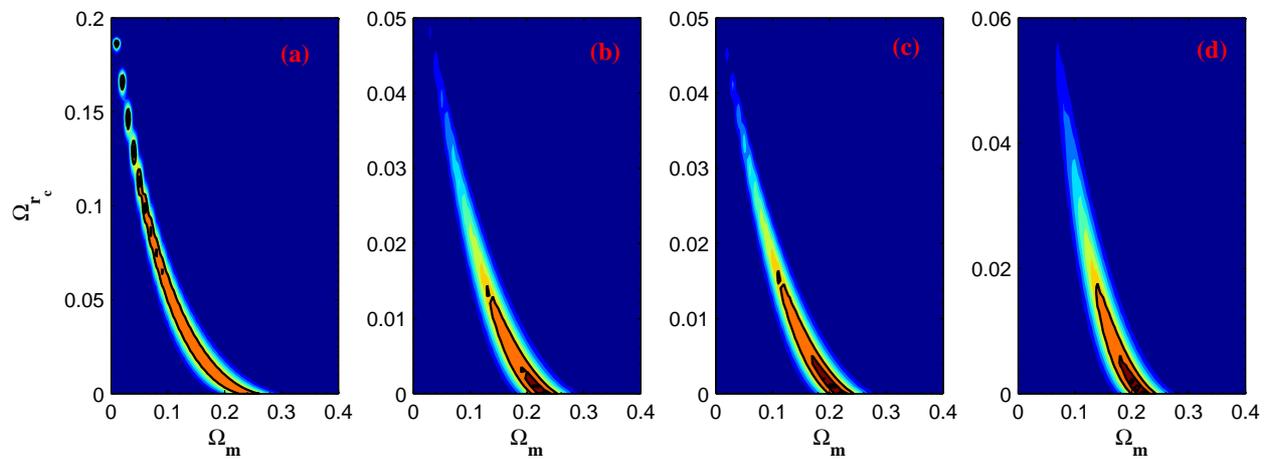} \vspace{4cm}
\end{center}
 \caption{\small {Contour maps of $\Omega_{r_{c}}$ versus $\Omega_{m}$
  for normal branch of WDGP model with $1\sigma \, (68.3\%)$ and
$2\sigma \, (95.5\%)$ confidence levels using SNIa (a), SNIa+CMB
(b), SNIa+CMB+SDSS (c) and SNIa+CMB+SDSS+Gas (d). We used the
Essence Sample for SNIa experiment..}}
\end{figure}
\noindent
\twocolumn

\subsection{The Age of the universe}

The age of the universe for an expanding universe in a flat
background is given by
\begin{equation}
t_{0}=\int_{0}^{t_{0}}dt=\int_{0}^{\infty}\frac{dz}{(1+z)H(z)}.
\end{equation}
The WMAP collaboration [13],\, using the $\Lambda CDM$ model,
quotes\, $t_{0} = 13.75\pm0.11$ Gyr. In our model $H(z)$ is given by
Eq. (9). A model-independent limit of $t_{0}$ can be obtained from
the age of the oldest stars, $t_{star}$. Since the age of the
universe should be greater than the age of the oldest stars, this
limit can be used as another probe to constrain our model ( see for
instance [19,20]). Studies on the old stars [21] suggest an age of
$13^{+4}_{-2}$ Gyr for the universe. Richer et. al. [22] and Hansen
et. al. [23] also proposed an age of $12.7\pm0.7$ Gyr, using the
white dwarf cooling sequence method. Recently, Frebel {\it et al.}
[24] have reported the discovery of HE $1523-0901$, with an age
$t_{star}=13.4\pm0.8(1\sigma)\pm1.8(syst)$ Gyr. With these
statistical and systematic errors, we adopt $t_{star}>12$ Gyr for
constraining our model [21]. Table $2$, shows that the age of the
universe is always greater than $t_{star}$ for all used data sets.
However, the SNIa data sets (Gold or Essence Samples) alone lead to
very great $t_{0}$ which is not reliable and it seems that the pure
SNIa test cannot give a good constraint on the model parameters
space. This feature is evident also in our previous relative
likelihood figures and has its origin in the \emph{geometric nature}
of the SNIa data sets. On the other hand the combined analysis of
SNIa with other data sets improves the constraints and gives a more
reasonable results.

\subsection{Effective Dynamics}
It has been shown in our previous work [15] that the Warped DGP
scenario has a phantom-like behavior without need to introduce a
phantom matter neither in the bulk nor on the brane. To investigate
the effective dynamics of the model we study the effective density
defined by
\begin{equation}
\rho_{eff}=\lambda+\frac{3\mu^{2}}{2r_{c}^{2}}\bigg[1-\sqrt{1+4r_{c}^{2}
\Big(\frac{\rho+\lambda}{3\mu^{2}}-\frac{{}\Lambda_{5}}{3}\Big)}\bigg]
\end{equation}
and an effective equation of state parameter
\begin{equation}
\omega_{eff}=-1-\frac{\dot{\rho}_{eff}}{3H\rho_{eff}}\,.
\end{equation}
So, the effective equation of state of dark energy can be expressed
as follows
\begin{footnotesize}
$$\omega_{eff}=-1+$$
\begin{equation}
\frac{-\Omega_{m}(1+z)^{3}}{\bigg(E^{2}-
\Omega_{m}(1+z)^{3}\bigg)\sqrt{1+\frac{4}{\Omega_{r_{c}}}\Big(\Omega_{m}(1+z)^{3}+
\Omega_{\lambda}+\Omega_{{}\Lambda_{5}}\Big)}}
\end{equation}
\end{footnotesize}
Using the best fit parameters obtained here and summarized in table
$2$,\, in figure $8$ we plot $\omega_{eff} $ versus the redshift for
combined SNIa+CMB+SDSS+galaxies clusters gas mass fraction data set
for Gold and Essence Samples.

\begin{figure}[htp]
\begin{center}\includegraphics{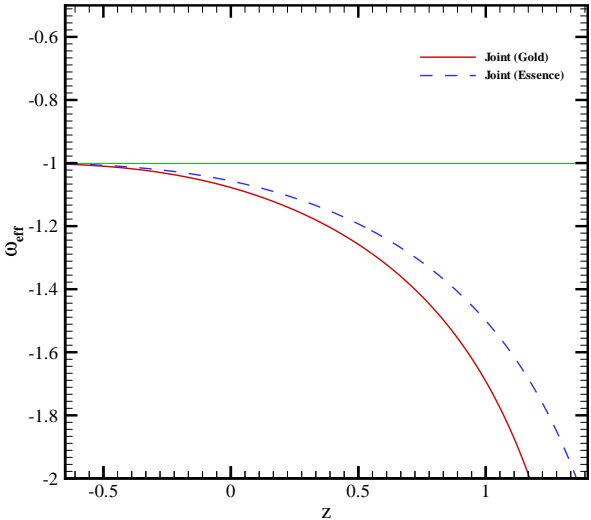} \vspace{6.5cm}
\end{center}
 \caption{\small {The effective equation of state parameter versus
the redshift fitted with SNIa+CMB+SDSS+galaxies clusters gas mass
fraction with Gold and Essence Samples respectively. }}
\end{figure}

As the figure shows, in low redshifts and especially at present,
dark energy exhibits a phantom-like acceleration with
$\omega_{eff}<-1$ ( it is easy to show also that in this model
$\dot{H}<0$ and $\dot{\rho}_{eff}>0$,\, see [15] for details). In
future ( $z<0 $) the effective equation of state parameter
approaches the cosmological constant line $\omega_{eff}=-1$. It is
important to note that although $\omega_{eff}$ has a phantom-like
dynamics in this warped DGP scenario, $\omega_{tot}$ doesn't show
such a behavior. The expression for $w_{tot}$ is given by
\begin{footnotesize}
$$\omega_{tot}=-1+$$
\begin{equation}
\frac{\Omega_{m}(1+z)^{3}}{E^{2}}
\Bigg(1-\sqrt{\frac{\Omega_{r_{c}}}{\Omega_{r_{c}}+\Omega_{m}(1+z)^{3}+
\Omega_{\lambda}+\Omega_{{}\Lambda_{5}}}}\Bigg).
\end{equation}\end{footnotesize}
Indeed, the total equation of state parameter remains
Quintessence-like ($w_{tot}>-1$) as figure $9$ shows.\\

\begin{figure}[htp]
\begin{center}\includegraphics{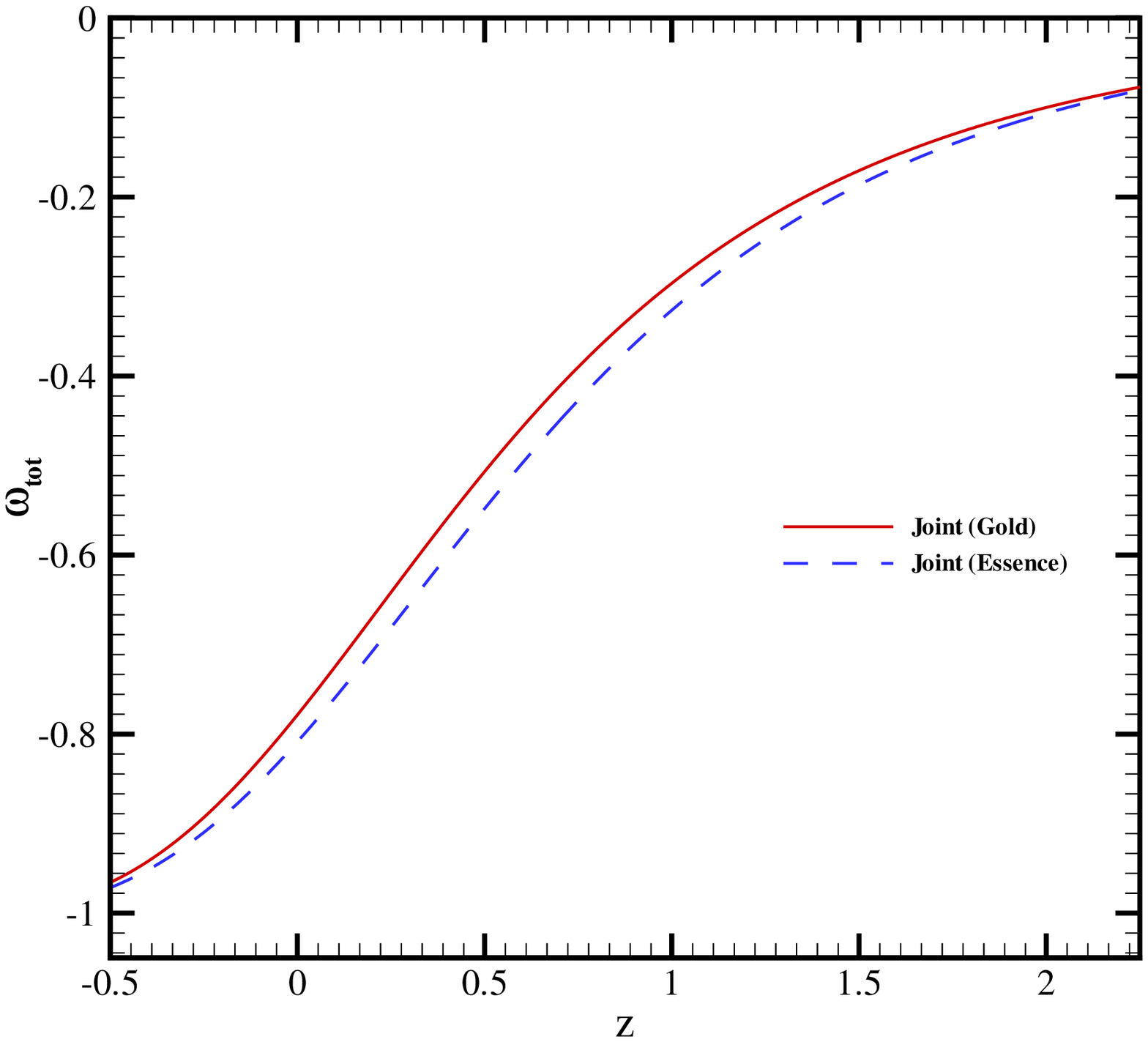} \vspace{6.5cm}
\end{center}
 \caption{\small {Total equation of state parameter versus
the redshift fitted with SNIa+CMB+SDSS+ galaxies clusters Gas mass
fraction with Gold and Essence Samples respectively. }}
\end{figure}

Another dynamical quantity of interest is the deceleration parameter
which is expressed as $q=-(1+\frac{\dot{E}}{H_{0}E^{2}})$\,where
\begin{footnotesize}
$$\frac{\dot{E}}{H_{0}}=\frac{-3}{2}\Big(\Omega_{m}(1+z)^{3}\Big)$$
\begin{equation}
\times\Bigg[1-\sqrt{\frac{\Omega_{r_{c}}}{\Omega_{r_{c}}+\Omega_{m}(1+z)^{3}+
\Omega_{\lambda}+\Omega_{{}\Lambda_{5}}}}\Bigg].
\end{equation}
\end{footnotesize}
This equation implies that for all values of $z$,\, $\dot{E}<0$. As
an important result, the deceleration parameter could be never less
than $-1$. Consequently, there is no super-acceleration or big rip
singularity in this model. Figure $10$ shows the variation of the
deceleration parameter versus the redshift with two combined data
sets. It is clear that the deceleration parameter reduces by
redshift towards recent epoch and in the future it approaches
$q=-1$.
\begin{figure}[htp]
\begin{center}\includegraphics{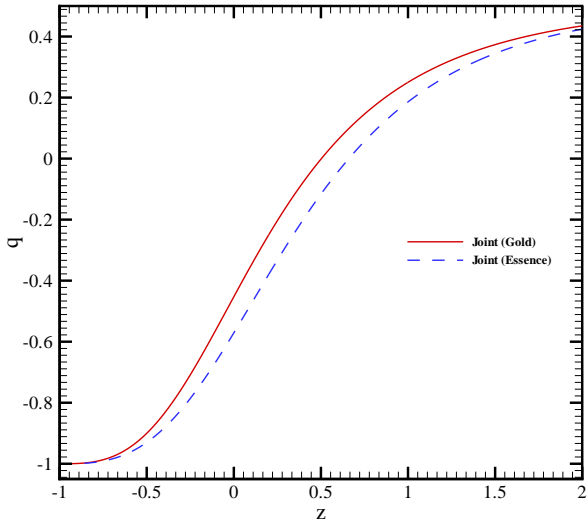}\vspace{6.5cm}
\end{center}
 \caption{\small {Variation of the deceleration parameter versus the
 redshift. The parameters values used to produce this figure are taken from
 table $2$. }}
\end{figure}

\section{Summary and Conclusion}
The normal branch of the warped DGP brane scenario has several
fascinating properties. Its cosmological dynamics does not suffer
from instabilities such as ghost, it can exhibit a phantom mimicry
and it leads to the famous $\Lambda$CDM scenario in certain limits.
In this regard, we have studied cosmological constraints imposed on
this model from observational data such as type Ia supernova data
from the Gold and Essence surveys, the baryon acoustic oscillations
(BAO) measurement from the Sloan Digital Sky Survey (SDSS), the
cosmic microwave background (CMB) and the baryon gas mass fraction
of clusters of galaxies. We find that the best fit values of model
free parameters are constrained to:
$\Omega_m=0.240^{+0.050}_{-0.130}$ and $\Omega_{r_c}=0.000^{+0.014}$
at $1\sigma$ confidence interval by using Gold sample SNIa $+$ CMB
shift parameter $+$ BAO $+$ gas mass fraction. The same analysis
just including essence sample SNIa instead of Gold sample correspond
to: $\Omega_m= 0.220^{+0.020}_{-0.170}$ and
$\Omega_{r_c}=0.000^{+0.025}$ at $68\%$ confidence interval.
Furthermore, using these best fit parameters, we obtained the age of
the universe $14.85 ^{+1.561}_{-3.942}$ and
$14.96^{+1.459}_{-7.888}$ at $1\sigma$ confidence interval by using
 SNIa$+$CMB shift parameter$+$BAO$+$gas mass
fraction for gold sample and essence sample of SNIa respectively. A
comparison between the age of the oldest stars and the one that we
have obtained by using the best fit values of parameters, show that
WDGP passes the age constraint very well. On the other hand we
investigated the effect of the WDGP model on the geometrical
parameters of the universe such as the transverse comoving distance
(comoving angular diameter distance), the comoving volume element
and the angular size. And finally, we studied the effective
cosmological dynamics of the model via effective equation of state
parameter and the deceleration parameter by using the best fit
parameters of the combined analysis of SNIa+CMB+SDSS+galaxies
clusters gas mass fraction with Gold and Essence Samples. This
analysis confirms that the WDGP model has a phantom-like behavior
without need to introduce phantom fields neither in the bulk nor on
the brane. It is important to note that in this scenario the effect
of the bulk's warped geometry (encoding in $\Lambda_{5}$) is that by
increasing the absolute value of the bulk cosmological constant the
phantom-like behavior reduces. In other words, incorporation of
$\Lambda_{5}$ leads to a reduction of the effective phantom nature
of the model in comparison with pure DGP case ( see Ref. [15]).

\bibliographystyle{elsarticle-harv}
\bibliography{<your-bib-database>}

\end{document}